\documentstyle[11pt,aaspp4]{article}

\def\lap{\hbox{${_{\displaystyle<}\atop^{\displaystyle\sim}}$}}

\slugcomment{MSUPHY98.17}

\begin{document}

\title{Low Frequency Gravitational Waves from Black Hole MACHO 
Binaries}

\author{William A. Hiscock}
\affil{Department of Physics, Montana State University, Bozeman,
Montana 59717}


\begin{abstract}
Nakamura, Sasaki, Tanaka, and Thorne have recently estimated the
initial distribution of binary MACHOs in the galactic halo assuming
that the MACHOs are primordial black holes of mass $\sim 0.5 M_\odot$,
and considered their coalescence as a possible source for ground-
based interferometer gravitational wave detectors such as LIGO.
Evolving their binary distribution forward in time to the present, the
low-frequency ($10^{-5} {\rm Hz}\, <\, f\, <\, 10^{-1} {\rm Hz}$)
spectrum of gravitational waves associated with such a population of
compact binaries is calculated.  The resulting gravitational waves
would form a strong stochastic background in proposed space
interferometers such as LISA and OMEGA.  Low frequency gravitational
waves are likely to become a key tool for determining the properties
of binaries within the dark MACHO population.  ( submitted to ApJ
Letters September 3, 1998)
\end{abstract}


\section{Introduction}

The MACHO project has thus far detected 15 candidate microlensing
events in the direction of the Large Magellanic Cloud (LMC)
(\cite{Cook98}) and 2 in the direction of the Small Magellanic Cloud
(SMC) (\cite{alco98}).  One of the LMC events (\cite{Bennett1996}) and
one of the SMC events (\cite{alco98}) appear to have been caused by
binary MACHO lenses.  The LMC observations suggest that perhaps half
of the halo consists of MACHOs of average mass $ \sim 0.5 M_\odot$.
Observational constraints make it unlikely that these MACHOs are main
sequence red dwarfs (\cite{bahc94}; \cite{graf96a};\cite{graf96b}), or
white dwarfs (\cite{char95}; \cite{adam96}).  Given these
difficulties, more speculative forms of MACHOs have been seriously
considered, such as primordial black holes or boson stars.

Recently, Nakamura, Sasaki, Tanaka, and Thorne (hereafter, NSTT)
considered the possibility that the halo MACHOs consist of primordial
black holes (BHMACHOs)(\cite{NSTT}).  They derived an estimate of the
initial binary distribution function, and determined the rate at which
BHMACHOs in binaries would coalesce due to emission of gravitational
radiation.  They found extragalactic halo BHMACHO coalescences to be
promising hypothetical sources of gravitational waves (GW) for ground
based interferometer detectors such as LIGO.

In this letter BHMACHOs are again examined as possible sources
of gravitational radiation, but now in the low frequency
($10^{-5} - 10^{-1} {\rm Hz}$) band where space-based interferometers,
such as the proposed LISA (\cite{LISA}) and OMEGA (\cite{OMEGA})
instruments will be most sensitive. The NSTT binary distribution
function is assumed, and evolved forward from the time of black
hole formation to the present, to determine the distribution of
BHMACHO binaries in semimajor axis and eccentricity today. The black 
hole binaries
with orbital frequencies higher than about $10^{-3.5}$ Hz are in
highly eccentric orbits, which greatly increases their efficiency as
sources of gravitational radiation, and spreads that radiation over
a large number of harmonics of the orbital frequency (\cite{PM1963}).
Binary BHMACHOs are shown to create a strong stochastic background 
in the low frequency GW band, possibly dominating over all other 
galactic background sources.

\section{Evolution of the NSTT binary distribution}

The distribution of black hole MACHO binaries obtained by
NSTT defines the probability distribution of values for the semimajor
axis $a$ and eccentricity $e$ of the binary's orbit. Utilizing a simple
model of primordial black hole formation, they found the 
probability distribution to be
\begin{equation}
F(a,e)\, da\, de \; = \; (3/2)a^{1/2}{\bar x}^{-3/2} e 
(1-e^2)^{-3/2}\, da\, de  ,
\label{Fae}
\end{equation}
where ${\bar x} \, = \, 1.1 \times 10^{16}\left({M_{BH}/M_\odot}\right)
^{1/3}(\Omega h^2)^{-{4/3}}\,\hbox{\rm cm}$,
and, as usual, $\Omega$ is the cosmological density parameter and $h$ is
the Hubble constant in units of $100\, {\rm km\, s^{-1}\, Mpc^{-1}}$.
Allowable values for the semimajor axis are bounded above by $\bar x$,
and the eccentricity for each value of $a$ is also bounded above, 
by $e_{\rm max} = [1-(a/{\bar x})^{3/2}]^{1/2}$. 

The spatial distribution of black hole MACHO binaries is assumed 
to follow the standard spherical flat rotation halo model  
\begin{equation}
	\rho \; = \; {\hat \rho}\, {R^2+a^2 \over r^2 + a^2} \; \; ,
\label{halo}
\end{equation}
where ${\hat \rho} = 0.0079 M_\odot\; {\rm pc}^{-3}$ is the local
density of dark matter, $r$ is the Galactocentric radius, $R = 8.5\;
{\rm kpc}$ is the Galactocentric radius of the Sun, and $a = 5.0 \;
{\rm kpc}$ is the halo core radius.  The distribution of
Eq.(\ref{halo}) is valid from $ r =0$ out to some halo radius, $R_h$;
typical values are $R_h = 50 \; {\rm kpc}$, for a halo extending to
the LMC; or $R_h = 300 \; {\rm kpc}$ for a halo extending half way to
M31.

The NSTT distribution given by Eq.(\ref{Fae}) is very strongly peaked
towards binaries with large eccentricities, due to the
$(1-e^2)^{-3/2}$ factor in $F$.  For example, for binaries with
initial orbital frequencies of $\sim 10^{-10}$ Hz (which will evolve
into the low frequency band today) 90\% of the binaries are within
$\delta e \simeq 0.00198$ of the maximum eccentricity, $e_{\rm max}
\simeq 0.9999775$.  In view of this, a simplifying assumption will be
made:  that all black hole MACHO binaries initially have the $e=e_{\rm
max}$.  The distribution of Eq.(\ref{Fae}) may then be rewritten in
simplified form, after an integration over $e$, as:
\begin{equation}
F(a) da\; =\; {3 \over 2} \left[ \left({a \over {\bar x}}\right)^{3/4} -
\left({a \over {\bar x}}\right)^{3/2} \right] {da \over a}
\label{Fai}
\end{equation}

The approximated NSTT distribution (with all binaries at their maximum
value of $e$) describes the binary black hole MACHO population
at the time of its formation, in the early universe. In order to 
calculate the spectrum of gravitational radiation produced by the
binary black holes, it is necessary to know the distribution of
binaries today. The population will have evolved due to a variety of 
effects, such as accretion, gravitational interactions with other 
MACHOs and with ordinary matter, and due to the emission of 
gravitational radiation. Only this last effect will be considered here.

In order to determine the probability distribution of black hole
MACHO binary orbital parameters today, the initial distribution 
given in Eq.(\ref{Fai}) must be evolved forward in time from the
early universe to the present epoch. The evolution of a binary 
system of point masses due to gravitational radiation reaction 
in the weak field limit is well understood (\cite {PM1963};
\cite{P1964}). The distribution was evolved by numerically
integration using a fourth-order adaptive stepsize 
Runge-Kutta routine. The results obtained are characterized by two
functions: $a_0(a_i)$, the present semimajor axis as a
function of the initial semimajor axis (a subscript $0$ indicates
the present day value of a quantity); and $e_0(a_i)$, the present
eccentricity as a function of the initial semimajor axis. 

The form of the distribution today is illustrated in Figure (1), in
which the present eccentricity, $e_0$, is plotted against the
present orbital frequency, $f_0$.
The circularizing effect of gravitational radiation reaction
is clearly visible; those binaries with orbital frequencies
today of $10^{-5}$ Hz or below are still highly eccentric,
while those which have evolved to higher frequencies such as
$10^{-2}$ Hz are basically circular today. Integration of a
$50 \; {\rm kpc}$ halo with $\Omega h^2 = 0.1$ gives $2.6 \times 10^7$
binaries in the low frequency band today, comparable to the
expected number of close white dwarf binaries in the Galaxy
(\cite{HBW}). In contrast, a $300 \; {\rm kpc}$ halo with
$\Omega h^2 = 1.0$ will have $ 1.6 \times 10^9$ black hole
binaries in the low frequency band today. 

\section{Stochastic gravitational wave background}

The calculation of the gravitational radiation spectrum emitted by
the halo population of BHMACHO binaries differs substantially
from the more familiar calculations for galactic white dwarf binaries.
This is because the BHMACHO binaries will generally have non-negligible
eccentricity. A binary system with an eccentric orbit will emit
significantly more power in GW than a circular system with the same
semimajor axis. Eccentric binaries will also
emit GW on a multitude of harmonics of the orbital frequency, whereas
circular binaries emit purely in the $n=2$ mode. Peters and Mathews
(\cite{PM1963}) found that the orbit-averaged power emitted in 
gravitational radiation by a binary system in the {\it n}th harmonic
of the orbital frequency is 
\begin{equation}
      L(f,n,e) = {32 \over 5} {m_1^2 \, m_2^2 \, (m_1 + m_2) \over a^5}
	g(n,e) \;\;,
\label{Lne}
\end{equation}
where $f$ is the orbital frequency associated with semimajor axis $a$ 
(not the frequency of the harmonic radiation), and $g(n,e)$ is given by
\begin{eqnarray}
	g(n,e)\; & =\; & {n^4 \over 32}  \bigg\{  \left[ J_{n-2}(ne)-
	2 e J_{n-1}(ne) + {2 \over n}J_n(ne) + 
	2 e J_{n+1}(ne) - J_{n+2}(ne) 
	\right]^2 \nonumber \\ & & +\left. \left(1-e^2 \right)
	\left[ J_{n-2}
	(ne)-2 J_{n}(ne)+J_{n+2}(ne) \right]^2  + {4 \over {3 n^2}}
	\left[ J_n(ne)\right]^2 \right\} \; \; .
\label{gne}
\end{eqnarray}

The total GW signal at frequency $f$ will incorporate contributions 
from a large number of binaries, and hence the signal will be 
stochastic in nature. The characteristic
spectral amplitude at frequency $f$ may be written as a sum over
the uncorrelated contributions from the binaries for which $f$ is
a harmonic,
\begin{equation}
        h_f(f) \; = \; {2 \over \pi f}  \left[\; \left\langle {1 \over
        d^2} \right\rangle \; \sum_{j=2}^{j_{\rm max}}\; L(f_{j},j,e_j)\;
        { dN \over df_{j}} \right]^{1/2} \; \; ,
\label{specamp}
\end{equation}
where $\langle d^{-2} \rangle$ is the inverse distance squared to the
source,
averaged over the source distribution, and the frequency $f_j$ is
the orbital frequency of a binary system for which the {\it j}th
harmonic is at frequency:  $f_j =f/j$. Similarly, 
$e_j$ is the eccentricity of the NSTT evolved binary for which 
the {\it j}th harmonic is at frequency $f$. The upper limit
on the summation, $j_{\rm max}$, is the value of $j$ for which the 
semimajor axis of the binary is at the
maximum value $a_i \sim {\bar x}$.   For each value of
$j$, $dN/df_j$ gives the number of sources per unit frequency interval
at frequency $f_j$, and $L(f_j,j)$ is the the power
emitted at frequency $f$ by a binary with orbital frequency $f_j$. 

The average squared inverse distance between halo sources and the Sun may
be found by integrating over the source distribution given in Eq.
(\ref{halo}). For a $50\; {\rm kpc}$ halo, this gives 
$\langle d^{-2} \rangle^{-1/2} = 16.8\;{\rm kpc}$,
while for a $300\; {\rm kpc}$ halo $\langle d^{-2} \rangle^{-1/2} = 39.8
\;{\rm kpc}$.

The procedure followed to determine the low frequency GW spectrum
of the NSTT distribution of black hole MACHO binaries is as follows.
First, a frequency $f$ is chosen. This fixes the frequencies (and
hence semimajor axes) of the binaries whose {\it j}th harmonic will
contribute to the sum in Eq.(\ref{specamp}).
For each harmonic index $j$ in Eq.(\ref{specamp}), the luminosity
$L(f_j,j)$ and the number of binaries per unit frequency interval,
$dN/df_j$, must then be determined.

In order to determine the
luminosity from Eq.(\ref{Lne}), the harmonic index $j$ must be
known, along with the semimajor axis, $a$, and the
eccentricity $e$. The semimajor axis is trivially obtained from
the orbital frequency $f_j$, while the eccentricity is also a
function of $f_j$, determined by the numerical evolution of the
approximate NSTT, as illustrated in Fig.(\ref{eccevol}).
Once the eccentricity is known, the function $g(j,e)$
may be evaluated. Direct numerical computation of Bessel function
values was used for the lowest ten harmonics ($j \leq 10$).
Accurate evaluation of the Bessel functions becomes more difficult for
larger values of $j$. An asymptotic expansion was
used to fix the value of $g(n,1)$ (which depends only on $J_n(n)$) and
the shape of $g(n,e)$ for arbitrary $e$ was then modeled
analytically using a combination of exponential and
trigonometric functions. Trial variation of parameters and 
functional forms within the analytic fit indicate that 
the resulting luminosity is robustly estimated to about 
one percent accuracy.

The evaluation of the number of binaries present today per unit
frequency interval, $dN/df$, for an arbitrary binary frequency $f$,
may be split into three pieces:
\begin{equation}
      {dN \over df} = N_{\rm halo}\; {dF \over da_i}\; {da_i \over da_0}
      \; {da_0 \over df} \; \; \; .
\label{dndf}
\end{equation}
Here $dF/da_i$ is the initial fraction of all black hole MACHO
binaries per unit semimajor axis interval, $da_i/da_0$
is the amount by which an infinitesimal interval in semimajor axis
has expanded or contracted in evolving to the present day, and
$da_0/df$ is the final conversion from semimajor axis to present
orbital frequency via Kepler's third law. Given an orbital
frequency $f_0$ today, the initial value of the semimajor axis $a_i$
may be determined from the results of the
numerical evolution of the approximate NSTT distribution. Then,
$dF/da_i$ may be directly evaluated from Eq.(\ref{Fai}).
The evolution of $da_i/da_0$
is also determined from the numerically evolved NSTT distribution.
Multiplication of the factors in Eq.(\ref{dndf}) gives the number
of binaries per unit frequency interval as a function of frequency,
today. The results confirm the notion that the total GW signal
from the black hole MACHO binaries will be stochastic in nature.
Assuming an integration time of 4 months (typical of LISA or OMEGA),
every $10^{-7}\, {\rm Hz}$ bin will include
modes from multiple binaries for all frequencies
$f \lap 10^{-2}\, {\rm Hz}$. Above that frequency, individual bins 
may be occupied by individual binaries evolving towards coalescence, or
may be empty, opening windows in which weaker sources (e.g.,
extragalactic) could be observed. 

The total gravitational wave spectral amplitude was then evaluated for
a set of frequencies, $f$, incorporating contributions from harmonics
up to a maximum value $j_{\rm max}$.  Setting $j_{\rm max}$ equal to
$1000$ was found to provide an accuracy over the entire frequency
range $10^{-5} - 10^{-1}\; {\rm Hz}$ of better than one percent;
specifically, the inclusion of $10^4$ harmonics rather than $10^3$
changes the spectral amplitude by less than 1\% over this range.

The resulting spectrum of gravitational radiation is shown in
Figure (\ref{fig2}). Both curves illustrated represent small halos,
$R_h = 50 \; {\rm kpc}$, with $\Omega h^2 = 0.1$ for the lower curve
and $\Omega h^2 = 1.0$ for the upper curve. The
spectrum is only weakly dependent of the value of
the halo radius; a larger halo simply adds additional sources
at large distance. Curves for a $300\; {\rm kpc}$ halo are of 
the same shape as those illustrated,
but with spectral amplitude increased by a constant factor of
$\sim 1.106$ over the full frequency range (on the logarithmic
plot of Fig(\ref{fig2}), this corresponds to moving the curves
upwards by  $0.0436$). 

The peak in the spectrum at around $f = 10^{-3}\; {\rm Hz}$ is due to
a combination of circumstances.  Binaries whose $n=2$ mode is at that
frequency have eccentricities of around $0.4$ (see
Fig.(\ref{eccevol})), so that still about 50\% of their power is being
emitted in the $n=2$ mode(\cite{PM1963}).  At the same time, lower
frequency binaries within an order of magnitude in frequency are
significantly eccentric, so that they contribute substantial power at
$10^{-3}\; {\rm Hz}$ through their higher harmonics.  At higher
frequencies, the lower frequency binaries are more nearly circular,
and their higher harmonics do not contribute so much power.  At lower
frequencies, the fundamental mode's contribution (and, later, the
higher modes' contributions) drops off sharply due to the $a^{-5}$
dependence in Eq.(\ref{Lne}).

At high frequencies, the spectrum is seen to approach a power
law behavior, $ h_f \sim f^{-7/6}$. Since binaries in the higher
frequency range are nearly circular, essentially all of the signal
is generated by the $n=2$ harmonic. The observed power law is
then explained by noting that $L \sim f^{10/3}$, while the factors
of $dN/df$ in Eq.(\ref{dndf}) have the following asymptotic
form: $dF/da_i \sim {\rm constant}, da_i/da_0 \sim f^{-2}, 
da_0/df \sim f^{-5/3}$. Using these asymptotic forms in Eq.(
\ref{specamp}) and only including the $j=2$ mode yields
$h_f \sim f^{-7/6}$.

Also shown in Fig.(\ref{fig2}) are the
sensitivity curves for the proposed LISA (\cite{LISA}) and OMEGA
(\cite{OMEGA}) space interferometer GW detectors, and a recent
estimate of the background due to galactic close
white dwarf binaries (CWDB) (\cite{HBPC}). Except at low frequencies,
the BHMACHO binary background would be notably stronger than the 
CWDB background. The halo black hole binaries would form a strong 
``confusion noise'' stochastic background, significantly larger
than the instrument noise in either proposed interferometer over
a large frequency range $10^{-4} {\rm Hz} < f < 10^{-2} {\rm Hz}$.
Of course, while the stochastic signal is in one sense a 
noise, it also provides direct information on the nature and 
distribution of MACHO binaries in the Galactic halo. 

Whether the halo MACHOs are primordial black holes, white dwarfs,
boson stars, or some other type of compact object, some fraction
of them will be in binary systems. Given the
huge number of MACHOs in the Galaxy's halo, the low frequency
gravitational wave signal from these binary systems will quite
likely be detectable by the space-based interferometers. These
interferometers will be excellent instruments for determining
the nature and population of the halo MACHO binaries. 

\acknowledgments
The author wishes to thank P. Bender, R. Hellings, and S. Larson 
for helpful discussions. This work was supported in part by 
National Science Foundation Grant No. PHY-9734834.

\newpage

\begin{figure}
\caption{The eccentricity $e_0$ of the evolved NSTT black hole MACHO
binaries is plotted against their orbital frequency $f_0$ today. } 
\label{eccevol} 
\end{figure}

\begin{figure} 
\caption{The spectral amplitude of gravitational waves
emitted by an evolved NSTT distribution of BHMACHO binaries in
a $50\; {\rm kpc}$ Galactic halo is shown for $\Omega h^2 = 0.1$ 
(lower curve) and $\Omega h^2 = 1.0$ (upper curve). Also shown are the
expected sensitivities of the proposed LISA and OMEGA space-based 
interferometers, and a recent estimate of the stochastic background due
to Galactic close white dwarf binaries (curve labeled CWDB).} 
\label{fig2} 
\end{figure}

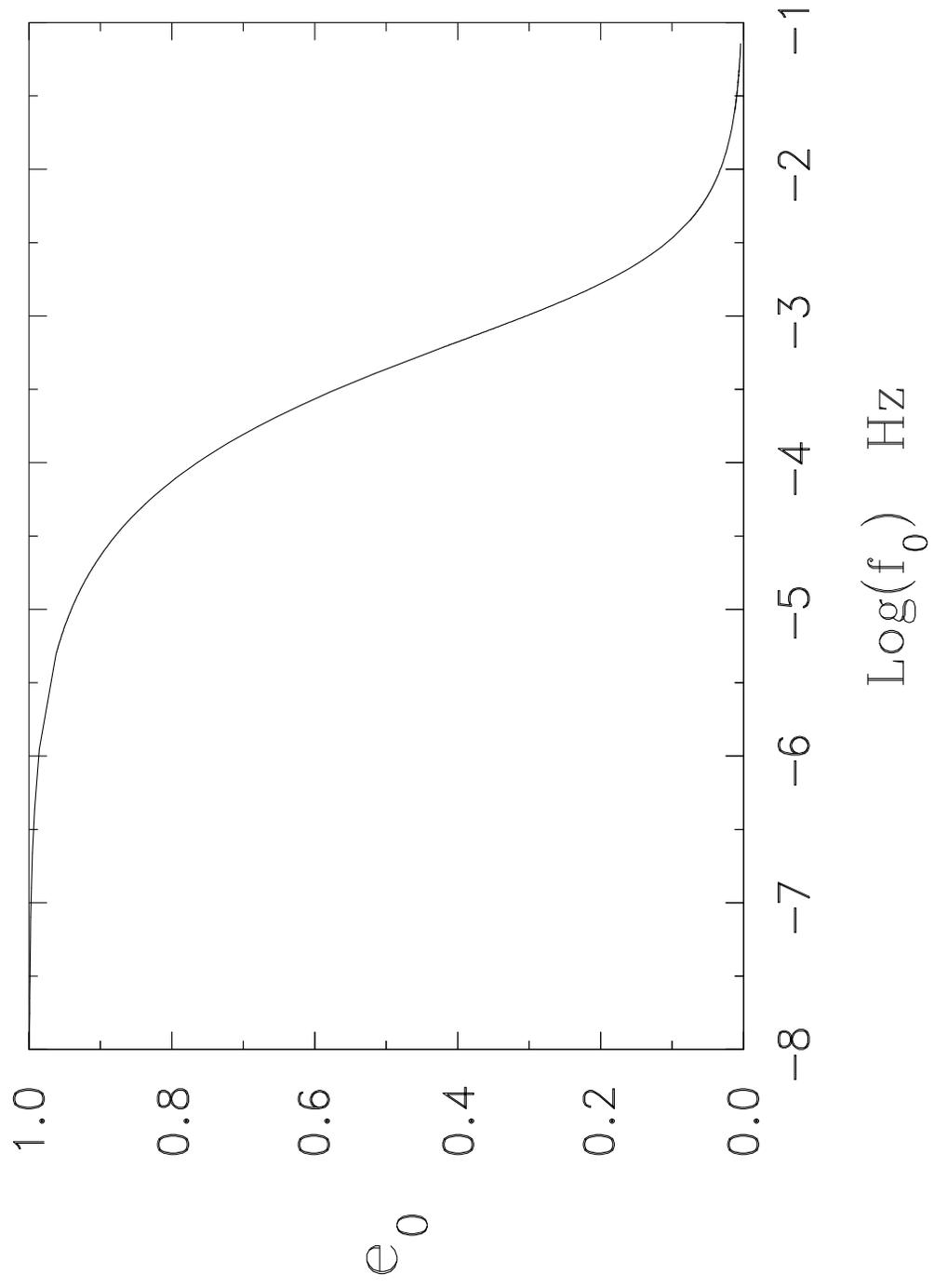
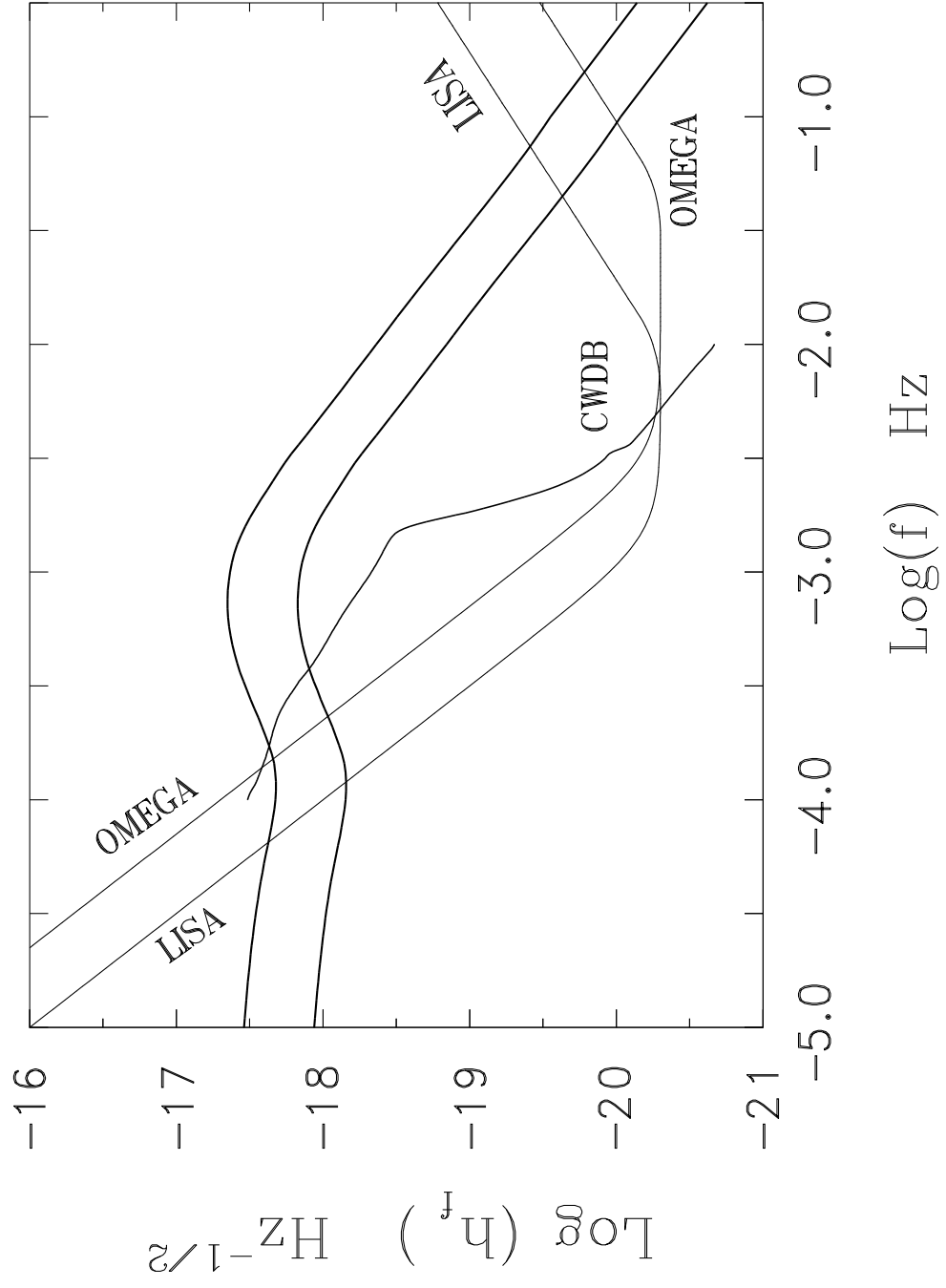
\end{document}